\documentclass[runningheads,citeauthoryear]{apinv}
\usepackage{epsfig,cite,graphics}
\usepackage{marvosym} %For astronomical symbols % % %
\usepackage[utf8]{inputenc}

\begin{document}

\title{Theoretical Model of Non-Conservative Mass Transfer with Uniform Mass Accretion Rate in Contact Binary Stars}
\titlerunning{Theoretical Model of Non-Conservative Mass}
\author{Prabir Gharami\inst{1},Koushik Ghosh\inst{2}, Farook Rahaman\inst{3}}
\authorrunning{P. Gharami, K.Ghosh, F.Rahaman}
\tocauthor{P. Gharami, K.Ghosh, F.Rahaman}
% Command tocautor{} is used by the Latex to give author names
% to the Contents of the volume (automatically generated)
\institute{Taki Bhabanath
High School, P.O.: Taki, North 24 Parganas, Pin-743429 West
Bengal, India .  e-mail: prabirgharami32@gmail.com
	\and Department of
Mathematics, University Institute of Technology, University of
Burdwan Golapbag (North), Burdwan-713104 West Bengal, India.\\  e-mail: koushikg123@yahoo.co.uk
	\and Department of
Mathematics, Jadavpur University, Kolkata 700032, West Bengal,
India. e-mail: rahaman@iucaa.ernet.in   \newline
	   }
\papertype{Submitted on xx.xx.xxxx; Accepted on xx.xx.xxxx}	
% Papertype can be "Research report", "Review", "Invited lecture", "Conference talk",
% "Conference poster", "Lecture at scientific seminar", "Summary of dissertation",  etc.
\maketitle

\begin{abstract}
In contact binaries mass transfer is usually non-conservative which
ends into loss of mass as well as angular momentum in the system. In
the present work we have presented a new mathematical model of the
non-conservative mass transfer with a uniform mass accretion rate in
a contact binary system with lower angular momentum. The model has
been developed under the consideration of reverse mass transfer
which may occur simultaneously with the original mass transfer as a
result of the large scale circulations encircling the entire donor
and a significant portion of the gainer. These circulations in
contact binaries with lower angular momentum are caused by the
overflow of the critical equipotential surface by both the
components of the binary system making the governing system more
intricate and uncertain.
\end{abstract}

\section{Introduction}

There exists a large number of binary systems in which the component
pairs of stars are with very small separations and with orbital
periods nearly less than 10 years such that mass can transfer from
one star to another. This mass transfer can change the structure of
both the stars causing subsequent evolution in the system as a whole
(Podsiadlowski  2001). These paired systems are known as contact
binaries. Thus mass transfer is a regular event in contact binary
stars. This mass transfer may occur in two ways. First type is
conservative in which total mass and angular momentum remains
unchanged. The second type is non-conservative where some mass is
lost during its journey from donor to gainer. Non-conservative mass
transfer is of two types: with uniform mass accretion rate
(Podsiadlowski et al.  1992; Sepinsky et al.  2006; Van Rensbergen
et al.  2010; Davis et al. 2013) and with non-uniform mass
accretion rate with respect to time (Stepien and Kiraga  2013;
Izzard et al.  2013; Gharami et al.  2015).

 In contact binaries
reverse mass transfer may occur simultaneously with this regular
mass transfer making a significant change in the overall dynamics
and evolution (Longair  1994; Nelson and Eggleton  2000; Stepien
2009; Stepien and Kiraga  2013). In case of contact binaries with
lower angular momentum the overflow of the critical equipotential
surface by both components drives large scale circulations
surrounding the entire donor and a huge fraction the gainer. A
portion of the mass transported by the donor to the gainer returns
back to the donor by this circulation with the mass flux of the
order of $10^{-3}$  to $10^{-4}$ $M_0$ per year (Stepien  2009).

In this paper we have  proposed a theoretical model of
non-conservative mass transfer with uniform mass exchange rate with
respect to time in contact binaries with lower angular momentum
under the consideration of reverse mass transfer following the
argument of Stepien  (Stepien  2009) . We have  furnished  a numerical model
for the presently proposed theory.

\section{              Theory }

We offer the following theoretical model of non-conservative mass
transfer with uniform mass accretion rate with respect to time in
contact binaries with lower angular momentum taking into account the
reverse mass transfer originated as a result of large scale
circulations encircling the entire donor and a major portion of the
gainer star. We here assume that $M_1$ is the mass of the gainer and
$M_2$ is the mass of the donor at any time t.

\begin{equation}
\dot{M_1}=\beta_0  \dot{M_2}^{(o)}
\end{equation}

\begin{equation}
  \dot{M_2}^{(o)}=-\frac{A}{\tau}M_2
\end{equation}

\begin{equation}
  \dot{M_2}=-\dot{M_2}^{(o)}+\dot{M_2}^{(i)}
\end{equation}

\begin{equation}
\dot{M_2}^{(i)}=\gamma (1-\beta_0) \dot{M_2}^{(o)}
\end{equation}

where $\beta_0$ characterizes the mass accretion process by the
gainer and $\gamma$ portrays the process of formation of reverse
mass jet directed to the donor.  (2) comes from the Bernoulli's law applied in the mass flow through the inner Lagrangian point considering adiabatic index   $\gamma =\frac{5}{3 } $  assuming the component stars being with convective envelops.
 A is a numerical constant lying
between 1 and 2 and $\tau$ is the entire timescale during which the
mass transfer is taking place (Pols  2012). $M_2^{(o)}$ and
$M_2^{(i)}$ indicate respectively the total  outgoing jet of mass
as a result of non-conservative mass transfer from donor to gainer
and total incoming mass flow directed towards the donor due to the
reverse mass transfer in time t.

Using (2) and (4) in (3) we get,

\begin{equation}
  \dot{M_2}=- [ 1- \gamma (1-\beta_0) ] \frac{A}{\tau}M_2
\end{equation}

Integrating equation (5) with the initial condition that at $ t=0,~
M_2=M_{2,0}$,  we get

\begin{equation}
  M_2 =M_{2,0}~e^{-\frac{A[ 1- \gamma (1-\beta_0) ]t}{\tau}}
\end{equation}

Again using (2) in (1) and on integration with the initial condition
that at $ t=0,~ M_1=M_{1,0}$  we have

\begin{equation}
  M_1 =M_{1,0}+\frac{\beta_0 M_{2,0}}{1- \gamma (1-\beta_0)} \left[ 1-e^{-\frac{A[ 1- \gamma (1-\beta_0)
  ]t}{\tau}} \right]
\end{equation}

Again using (6) in (2) we get,

\begin{equation}
\dot{M_2}^{(o)}=\frac{A}{\tau}M_{2,0}~e^{-\frac{A[ 1- \gamma
(1-\beta_0) ]t}{\tau}}
\end{equation}

Integrating (8) with the initial condition $M_2^{(o)}=M_{2,0}^{(o)}$
at $t=0$,   we get

\begin{equation}
M_2^{(o)}=M_{2,0}^{(o)} +\frac{  M_{2,0}} {1- \gamma
(1-\beta_0)} \left[ 1-e^{-\frac{A[ 1- \gamma (1-\beta_0)
  ]t}{\tau}} \right]
\end{equation}

Again using (2),  we get from (4)

\begin{equation}
\dot{M_2}^{(i)}= \gamma (1-\beta_0) \frac{A}{\tau}M_2
\end{equation}

Integrating equation (10) with the initial condition at $ t=0,~
M_{2,0}^{(i)}=0$  (understandably there should not be any initial
jet of reverse flow at the very beginning instant of mass transfer)
we get,

\begin{equation}
M_2^{(i)}=\frac{ \gamma (1-\beta_0)}{1- \gamma (1-\beta_0)}M_{2,0}
\left[ 1-e^{-\frac{A[ 1- \gamma (1-\beta_0)
  ]t}{\tau}} \right]
\end{equation}

\section{\textbf{Results:}}

Here we produce a numerical example   taking  the initial masses of
the gainer and donor as $M_{1,0}=9\times10^{31} $(g) and
$M_{2,0}=4\times10^{32 }$(g) respectively. For the present calculation we take
the values of the parameters as $A=1.5$, $\gamma =0.1$ and  $\beta_0
=0.4$. The time scale of the mass exchange is taken as  $\tau=10^4$ ( years) and
the outgoing mass from the donor at initial instant i.e. at time
$t=0 $ is taken as $M_{2,0}^{(o)} = 1.0 \times 10^{29}$ (g). Also
$M_{2,0}^{(i)} =0 $ i.e. at the initial instant no incoming mass jet
is taken into account towards the donor. The orbital period of the system is considered to be less than 100 days so that the gainer can fill the Roche lobe during its expansion (Monzoori 2011).
Here we have produced the graphs for mass incoming to the donor vs. time (years) (Figure 1), mass outgoing from the donor vs. time (years) (Figure 2), mass of the donor vs. time (years) (Figure 3) and mass of the gainer vs. time (years) (Figure 4). Figure 1, 2 and 4 show increasing profiles while Figure 2 for obvious reason exhibit decreasing profile.

\pagebreak

\section{Discussion:}

The present work proposes a theoretical model of non-conservative mass transfer with uniform mass accretion rate with respect to time in contact binaries with lower angular momentum under the consideration of reverse mass transfer based on the proposal of Stepien (2009) which points out a possible large scale circulation generated by the overflow of the critical equipotential surface by both components as the possible driving force of this reverse mass transfer. Stepien and Kiraga (2013) proposed an alternative cause of reverse mss transfer in contact binaries. They proposed that as the altitude of the equatorial bulge quantifies to a certain percentage of the stellar radius above the surface hence when the radius of the accretor tends to the size of the Roche lobe by this quantity the apex of the bulge starts to protrude beyond the inner critical surface. As a result, part of the matter flows above the Roche lobe and comes back to the donor.  In future, we may work in this aspect of reverse mass transfer. Moreover, the present work focuses only on the issue of non-conservative mass transfer with uniform accretion rate. Future work may be carried in the direction of non-uniform mass accretion rate. There is also a provision to study the effect of magnetic field in reverse mass transfer. Theoretical study in future in this direction should bring some interesting results.
\begin{figure}
    \centering
        \includegraphics[scale=.5]{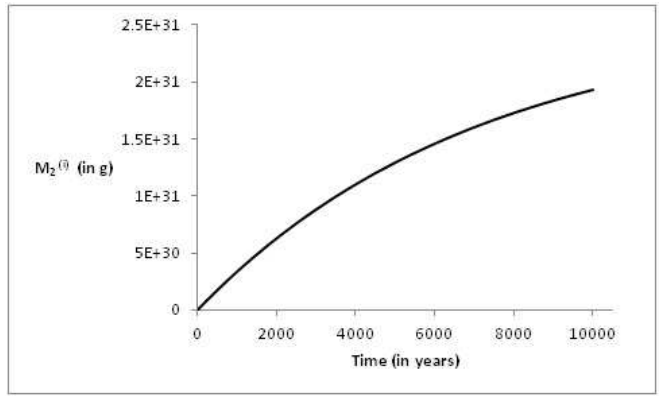}
    \caption{Mass incoming to the donor.}
    \label{Fig1}
\end{figure}
\begin{figure}
    \centering
        \includegraphics[scale=.5]{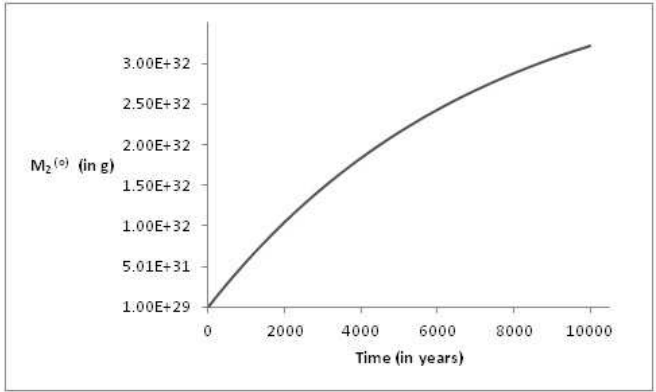}
    \caption{Mass outgoing from the donor . }
    \label{Fig2}
\end{figure}
\begin{figure}
    \centering
        \includegraphics[scale=.5]{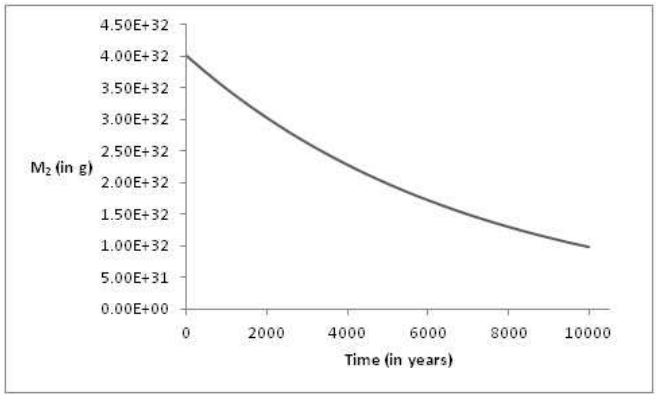}
    \caption{ Mass of the donor . }
    \label{Fig3}
\end{figure}
\begin{figure}
    \centering
        \includegraphics[scale=.5]{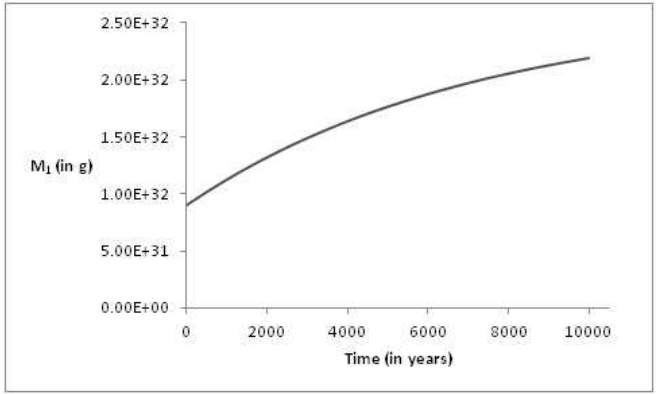}
    \caption{Mass of the gainer.}
    \label{Fig4}
\end{figure}

\pagebreak

\section{ Acknowledgments}
 FR gratefully acknowledges support from the Inter-University Centre for Astronomy and Astrophysics (IUCAA), Pune, India for providing research facility. PG, KG and FR express their sincere gratitude to Prof. Probhas Raychaudhuri, Professor (Retd.), Department of Applied Mathematics, University of Calcutta, Kolkata, India for a careful reading of the manuscript and for his valuable suggestions in order to improve the quality of the present work.

\begin{thebibliography}{99}

1.   	Davis, P.J., Siess, L. $\&$ Deschamps, R. 2013, A $\&$ A, 556, 4.\\

2.	Gharami, P., Ghosh, K. $\&$  Rahaman, F. 2015, arXiv:1407.2498 [gr-qc].\\

3.	Izzard, R.G., de Mink, S.E., Pols, O. R., Langer, N., Sana, H. $\&$ de Koter, A. 2013, Mem.S.A.It., 1, 1.\\

4.	Longair, M.S. 1994, Vol.2, Cambridge University Press, 41.\\

5.	Monzoori, D. 2011, in Advanced Topics in Mass Transfer, (ed.) by El-Amin, M., In Tech., 163.\\

6.	Nelson, C.A. $\&$ Eggleton, P.P. 2001, Astrophys. J., 552, 664.\\

7.	Podsiadlowski, P. 2001, in Accretion Processes in Astrophysics, (ed.) Matrtinez-Pais, I.G., Shahbaz, T. and Velazquez, J.C., Cambridge University Press, 45.\\

8.	Podsiadlowski, P., Joss, P.C. $\&$ Hsu, J.J.L. 1992, Astrophys. J., 391, 246.\\

9.	Pols, O. R. 2012, Lecture notes on Binary Stars, Department of Astrophysics/IMAPP, Radboud University Nijmegen, The Netherlands.\\

10.	Sepinsky, J.F., Willems, B., Kalogera, V. $\&$ Rasio, F.A. 2006, arxiv: 0903.0621v1.\\

11.	Stepien, K.  2009, MNRAS, 397, 857.\\

12.	Stepien, K. $\&$ Kiraga, M. 2013,  Acta Astronomica, 63, 239.\\

13.	Van Rensbergen, W., De Greve, J.P., Mennekens, N., Jansen, K. $\&$ De Loore, C. 2010, A $\&$ A, 510, A13.

\end {thebibliography}

\end{document}